# High Breakdown Electric Field (> 5 MV/cm) in UWBG AlGaN Transistors

Seungheon Shin[1,a)], Hridibrata Pal[3], Jon Pratt[1], John Niroula[3], Yinxuan Zhu[1], Chandan Joishi[1], Brianna A. Klein[4], Andrew Armstrong[4], Andrew A. Allerman[4], Tomás Palacios[3], Siddharth Rajan[1,2,a)]

[1]*Department of Electrical & Computer Engineering, The Ohio State University, Columbus OH 43210, USA*
[2]*Department of Materials Science & Engineering, The Ohio State University, Columbus OH 43210*
[3]*Microsystems Technology Laboratories, Massachusetts Institute of Technology, Cambridge, Massachusetts 02139, USA*
[4]*Sandia National Laboratories, Albuquerque, New Mexico 87123, USA*

**Abstract:**

**We report on the design and demonstration of ultra-wide bandgap (UWBG) AlGaN-channel metal-insulator heterostructure field effect transistors (HEFTs) for high-power, high-frequency applications. We find that the integration of gate dielectrics and field plates greatly improves the breakdown field in these devices, with state-of-art average breakdown field of 5.3 MV/cm (breakdown voltage > 260 V) with an associated maximum current density of 342 mA/mm, and cut-off frequency of 9.1 GHz. Furthermore, low trap-related impact was observed from minimal gate and drain lag estimated from pulsed I-V characteristics. The reported results provide the potential of UWBG AlGaN HEFTs for the next generation high-power radio frequency applications.**

[a)] Authors to whom correspondence should be addressed

Electronic mail: *shin.928@osu.edu, rajan.21@osu.edu*

Ultra-wide bandgap (UWBG) AlGaN is a promising candidate material for future RF and mm-wave applications due to its high breakdown field and excellent transport properties. The Johnson figure of merit (JFOM)—a key figure of merit to estimate the product of the cut-off frequency ($f_T$) and breakdown voltage ($V_{BR}$) – is expected to be 22 THz·V [1], which is significantly higher than that of state-of-art GaN or InP devices. However, high contact resistance remains a practical limitation to AlGaN transistors. To improve ohmic contact properties, recent work by Zhu et al. achieved a state-of-art $R_C$ of 0.25 Ω·mm to n-type AlGaN by incorporating reverse-graded contacts and bandgap narrowing effect in UWBG AlGaN [7], while approaches such as advanced regrowth techniques have also shown encouraging results [14, 33, 34]. Several demonstrations of UWBG AlGaN heterostructure field effect transistors (HEFTs) over the last decade have shown consistent improvement [9]. However, the reported average breakdown fields ($F_{BR}$) in AlGaN transistors have been relatively low, typically below 3.6 MV/cm. [16]. In this study, we report on the successful demonstration of UWBG AlGaN metal-insulator HFETs with breakdown fields in excess >of 5 MV/cm.

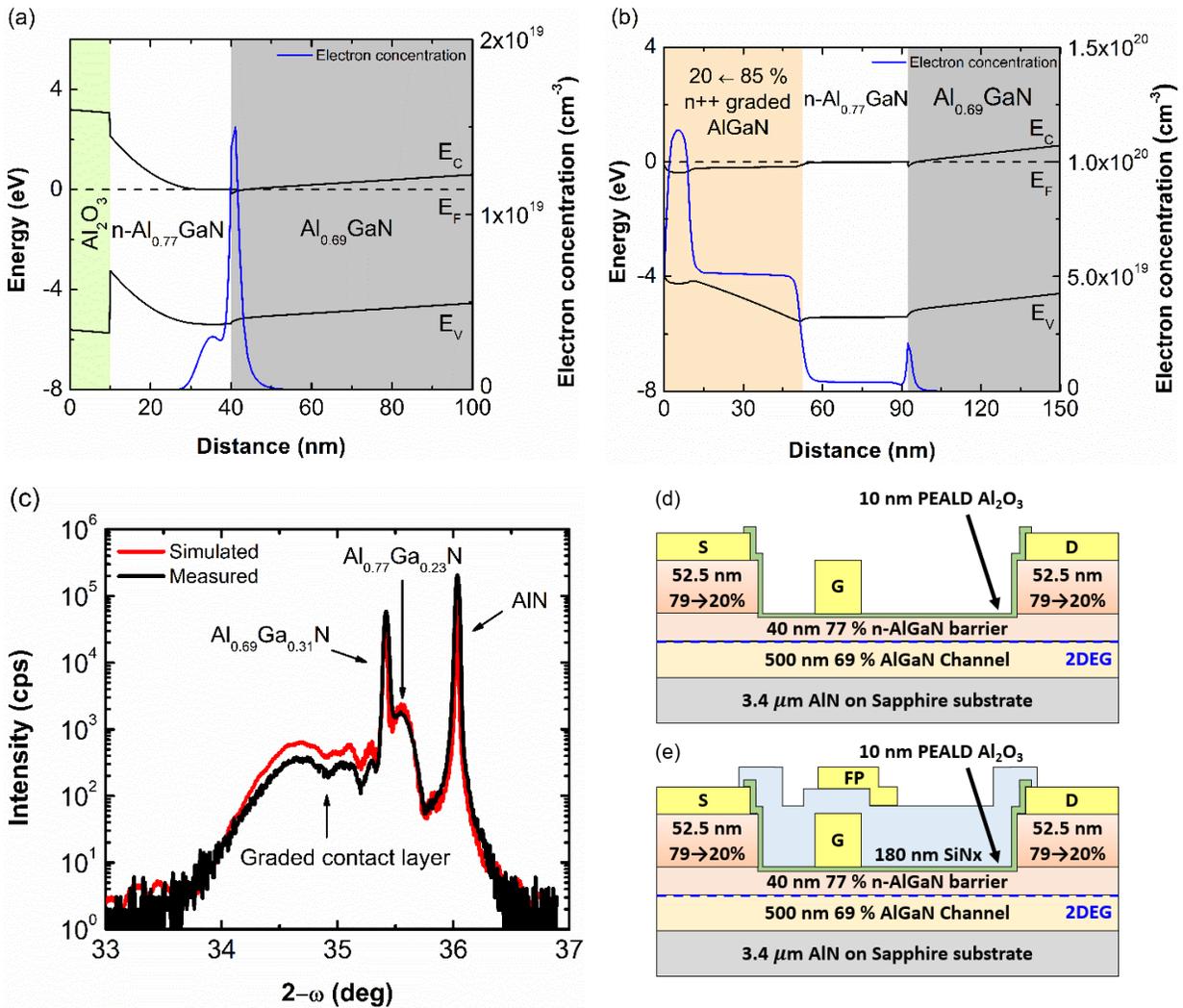

Figure 1. Calculated equilibrium energy band diagram (a) under the gate, (b) under the ohmic contact region, (c) XRD 2θ-ω scan of epitaxial stack, Schematic of device structures (d) Sample B, (e) Sample C

The epitaxial layers used for device fabrication were grown on a TNSC-4000HT metal-organic chemical vapor deposition (MOCVD) reactor on previously grown AlN/Sapphire templates. The epitaxial stack consists of 500 nm 69% Al-content AlGaN channel, a 40 nm thick 77 % n-AlGaN barrier (Si = 4 × $10^{18}$ cm$^{-3}$), and reverse-graded contact layers (Al% ~ 79 % to 20 %) [7]. The 40 nm-thick barrier was designed to enable over-etching into the barrier, ensuring the complete removal of the contact layer in the access region after etching, while also preventing ICP-RIE-induced damage. The AlGaN layer Al compositions and thicknesses were confirmed with high-resolution x-ray diffraction (HR-XRD) measurements on a Bruker D8 Discover system. As shown in Figure 1(c), the measured and simulated XRD spectra show good match.

Figure 1(a), (b) presents the equilibrium energy band diagram for gate (Figure. 1(a)) and ohmic region (Figure 1(b)), which were built based on the XRD results. The barrier height between Ni/Al$_2$O$_3$ was assumed as 3.2 eV [35-36]. From the simulated energy band diagram, the expected total charge density and two-dimensional electron charge density (n$_{2DEG}$) were 7.16 × $10^{12}$ cm$^{-2}$ and = 4.72 × $10^{12}$ cm$^{-2}$, respectively.

Table 1. Device Dimensions

|  | $L_{SG}$ (μm) | $L_G$ (μm) | $L_{GD}$ (μm) | $t_{ox}$ | Passivation | $L_{FP}$ (μm) |
|---|---|---|---|---|---|---|
| Sample A | 0.5 | 0.8 | 0.7 | - | - | - |
| Sample B | 0.5 | 1 | 0.5 | 10 nm Al$_2$O$_3$ | - | - |
| Sample C | 0.5 | 1 | 0.5 | 10 nm Al$_2$O$_3$ | 180 nm SiN$_x$ | 0.25 |

Direct-write optical lithography was used for patterning of all deposition and etch steps described below. The contact layer was etched using a low-damage ICP-RIE process (BCl$_3$/Cl$_2$/Ar = 5/50/5 sccm, ICP/RIE power = 40/8 W at 5 mTorr) into the 77 % barrier layer. The estimated barrier thickness after etching was estimated to be 29.5 nm using atomic force microscopy. Ti/Al/Ni/Au (20/120/30/100 nm) ohmic metal was patterned using electron-beam evaporation and lift-off in NMP. A 10 nm PEALD Al$_2$O$_3$ was deposited at 250 °C using TMA and O$_2$ as precursors with a growth rate of 1 Å/cycle. Mesa isolation was done using ICP-RIE etching (etch depth of 200 nm) and Ni/Au/Ni (30/100/50 nm) gate layers were deposited using e-beam evaporation and lift-off. Plasma-enhanced chemical vapor deposition (PECVD) SiN$_x$ (180 nm) was deposited as a passivation layer, and Ni/Au gate-connected field plates were evaporated. The electrical characteristics of three samples (Table 1) are outlined in this paper: (1) Sample A, field effect transistors fabricated without Al$_2$O$_3$ gate dielectric SiN$_x$, (2) Sample B, transistors with Al$_2$O$_3$ gate dielectric but with no field plate, and (3) and Sample C, transistors with Al$_2$O$_3$ gate dielectric, SiN$_x$ passivation and a field plate (Figure 1(b)-(c)).

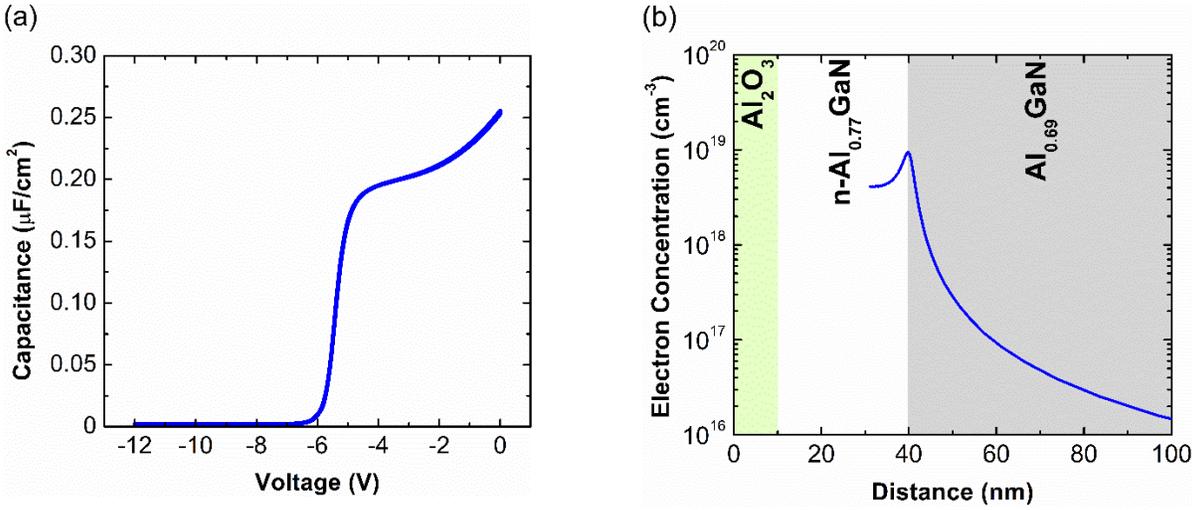

Figure 2. (a) Double-sweep C-V results after $Al_2O_3$ integration, (b) extracted electron concentration profile corresponding to gate sweep from 0 V to -12 V.

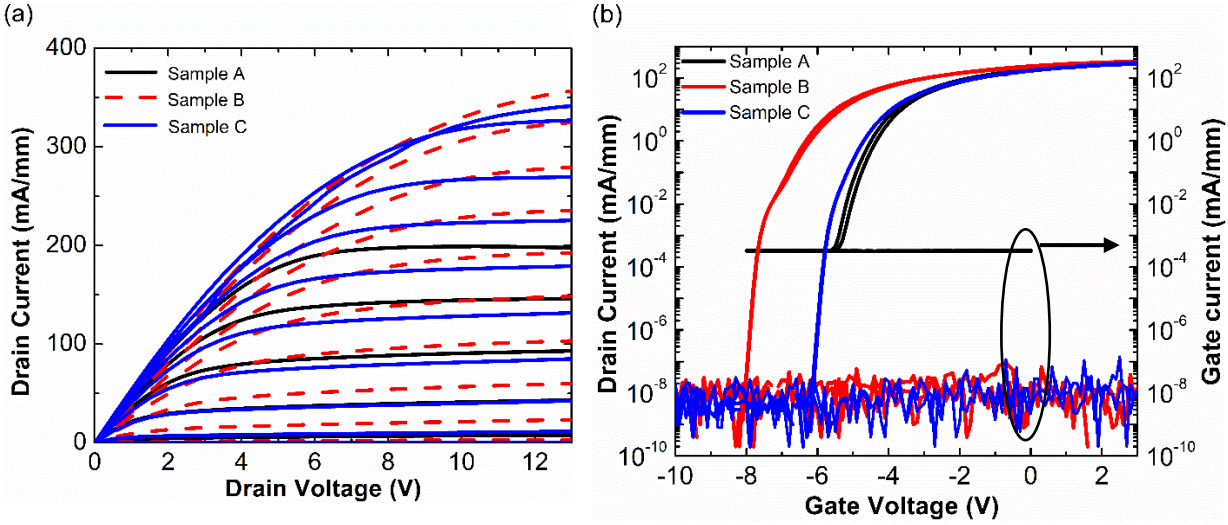

Figure 3. (a) Output curves with DC bias condition of $\Delta V_{GS}$ = -1 V, $V_{GS}$ = 0 ~ -5 V for Sample A, $\Delta V_{GS}$ = -1 V, $V_{GS}$ = 3 ~ -6 V for Sample B and C, (b) transfer curves at $V_{DS}$ = 10 V for different device structures

The sheet resistance, Hall carrier density, and Hall mobility for the channel regions (with the contact layer etched away) were estimated to be 6.79 $k\Omega/\square$, $6 \times 10^{12}$ $cm^{-2}$, and 153 $cm^2/V \cdot s$, respectively. A contact resistance of 3.26 $\Omega \cdot mm$ and specific contact resistivity of $1.86 \times 10^{-5}$ $\Omega \cdot cm^2$ were estimated from transmission line measurements (TLM). The contact resistance is higher than that seen for reverse-graded contacts made directly to n-type AlGaN [7]. We attribute this to the additional conduction band barrier introduced by the heterojunction. Capacitance-voltage (C-V) and DC I-V characteristics were measured using a Keysight B1500A. C-V measurements on metal-$Al_2O_3$-AlGaN capacitors showed hysteresis-free behavior under double-sweep voltage measurements (Figure 2(a)), and the integrated sheet charge density from pinch-off to zero bias was estimated to be $7.15 \times 10^{12}$ cm$^{-2}$ (Figure 2(b)). The charge profile extracted from C-V measurements suggests apparent charge density ($\sim 4 \times 10^{18}$ cm$^{-3}$) in the barrier layer, and a 2-dimensional electron gas as expected at the heterojunction, at a depth of 39.5 nm. Integration of the charge profile shows that the majority of the charge ($4.6 \times 10^{12}$ cm$^{-2}$) is contained in the 2-dimensional electron gas.

Sample B showed threshold voltage ($V_{TH}$) of -8 V and maximum on-current ($I_{MAX}$) of 358 mA/mm at $V_{GS}$ = +3 V (Figure 3(a)), while Sample C exhibited $V_{TH}$ = -7 V, and $I_{MAX}$ = 342 mA/mm at $V_{GS}$ = 3 V and. We attribute the relatively minor $V_{TH}$ variation between samples to differences in the epitaxial layer structure. At $V_{DS}$ = 10 V the maximum transconductance ($g_{m.MAX}$) for Sample A was 48 mS/mm and that for Sample B was 49 mS/mm. The gate leakage current for the baseline HFET was relatively high without the $Al_2O_3$ gate dielectric, resulting in $I_{ON}/I_{OFF}$ ratio ($< 6 \times 10^5$) (Figure 3(b)). However, integration of 10 nm $Al_2O_3$ significantly reduced the gate leakage current to sub nA/mm levels, enhancing the $I_{ON}/I_{OFF}$ ratio to $2 \times 10^{10}$.

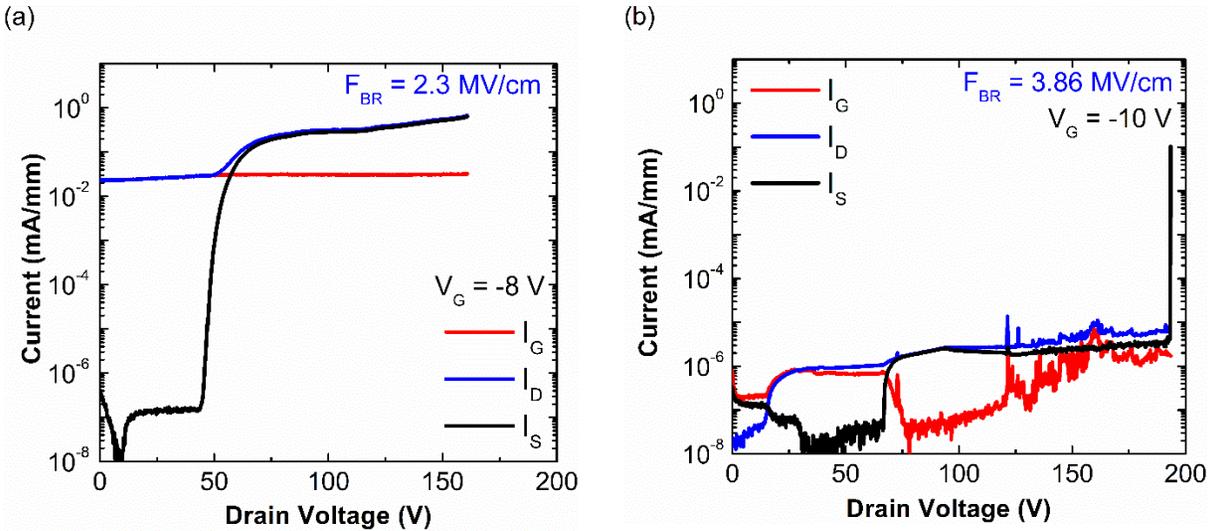

Figure 4. Three-terminal currents as a function of drain bias for (a) Sample A, and (b) Sample B

Three-terminal off-state leakage currents were measured using Keysight B1500A within the 200 V DC bias range while higher voltage characteristics were investigated using a Keysight B1505A power device analyzer. Without $Al_2O_3$ gate dielectric, relatively high gate leakage was observed, possibly due to the high doping concentration in the AlGaN barrier layer (Figure 4(a)), but the leakage current reduces after deposition of $Al_2O_3$, as shown in Figure 4(b). For Sample B, the gate leakage remained below $1 \times 10^{-5}$ mA/mm at $V_{GS}$ = -10 V and $V_{DS}$ = 193 V (Figure 4(b)). This result is consistent with previous work on

PEALD $Al_2O_3$/AlGaN structures, and shows that the dielectric effectively reduces and suppresses gate leakage current [10].

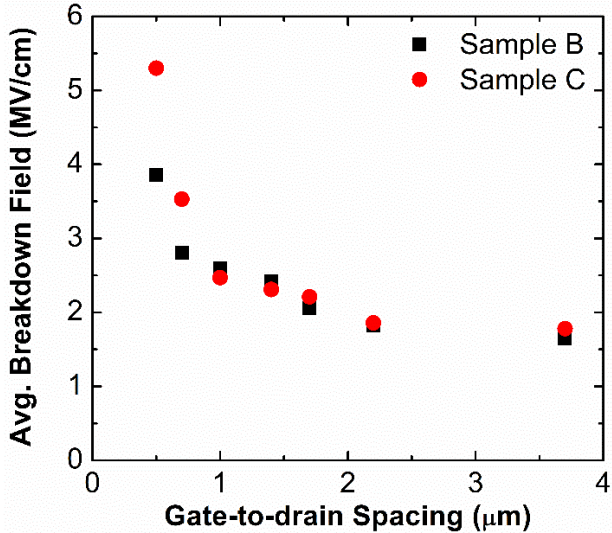

Figure 5. Average maximum breakdown field ($F_{BR}$) as a function of the gate-drain spacing ($L_{GD}$) transistors with gate dielectric/no passivation (Sample B), and with gate dielectric + passivation + field plate (Sample C)

The $L_{GD}$-dependent relationship between $F_{BR}$ and breakdown voltage was investigated by measuring the breakdown voltage while varying the gate-to-drain spacing. By plotting $F_{BR}$ as a function of $L_{GD}$, it was observed that $F_{BR}$ decreases with increasing $L_{GD}$ (Figure 5). The breakdown voltage for the purpose of this paper was defined as the voltage where the current reaches 1 mA/mm. For Sample B, a maximum $V_{BR}$ 193 V was observed at $V_{GS}$ = -10 V, corresponding to an $F_{BR}$ of 3.86 MV/cm, with $V_{BR}$ variations ranging from 150 to 193 V depending on the device position. Breakdown drain voltage of 260 V at a gate bias of -11 V was measured for Sample C (includes $SiN_x$ passivation and the gate-connected field plate) (Supplementary Figure S1 (b)). The highest average field ($F_{BR}$) is therefore 5.3 MV/cm, not including the gate bias potential itself. The breakdown voltage varied between 185 V and 260 V across the wafer. To the best of our knowledge, these off-state transistor breakdown voltage values represent the current state-of-art among reported UWBG AlGaN transistors, and are very similar to the highest breakdown field seen in another ultra-wide bandgap semiconductor [38, 39]. It is encouraging to see significantly higher breakdown field when compared with typical GaN-channel devices ($F_{BR}$ ~ 1-2 MV/cm), and confirms the potential of UWBG AlGaN for high-power RF applications [11].

At longer gate-drain distances, the breakdown voltage increased but at a slower rate, leading to lower average breakdown electric field. For example, at an $L_{GD}$ = 3.7 $\mu$m a breakdown voltage of 660 V was measured (Supplementary Figure S1 (b)). These excellent off-state characteristics point to the high intrinsic breakdown strength of the barrier and channel AlGaN material, as well as the high quality of the dielectric $Al_2O_3/Al_{0.77}GaN$ interface [10]. This result supports the potential of highly scaled UWBG AlGaN transistors for THz-frequency switching applications with enhanced breakdown performance.

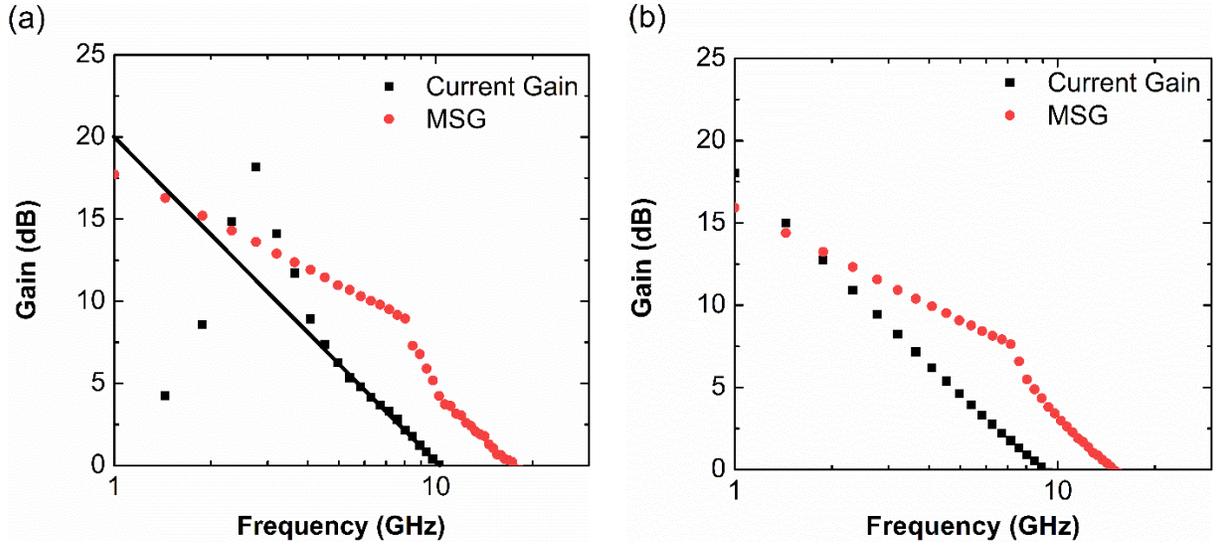

Figure 6. Small signal measurements for (a) Sample B measured at $V_{GS}$ = -3 V and $V_{DS}$ = 13 V, exhibiting $f_T$ = 10.3 GHz, $f_{MAX}$ = 17.5 GHz (b) Sample C at $V_{GS}$ = -2.5 V and $V_{DS}$ = 10 V, showing $f_T$ = 9.1 GHz, $f_{MAX}$ = 15 GHz

On-wafer small-signal measurements were performed on devices with dimensions similar to those used for on-state and off-state DC characterization. The measurements were done using Agilent vector network analyzer 8510C. The DC bias point used for Sample B was $V_{GS}$ = -3 V and $V_{DS}$ = 13 V, and $V_{GS}$ = -2.5 V and $V_{DS}$ = 10 V for Sample C. Figure 6 displays the estimated short-circuit current gain and unilateral power gain in dB scale as a function of measured frequency. Sample B had 10.3 GHz of $f_T$ and 17.5 GHz of $f_{MAX}$, whereas Sample C exhibited 9.1 GHz of $f_T$ and 15 GHz of $f_{MAX}$. We suspect decreased RF performances of Sample C may possibly be due to the additional parasitic capacitance components induced by the device structure and field plate with passivation. Although the RF characteristics exhibited a slight degradation due to the field-plated structure, this approach remains a viable electric field management technique, considering the significant improvements in breakdown performance

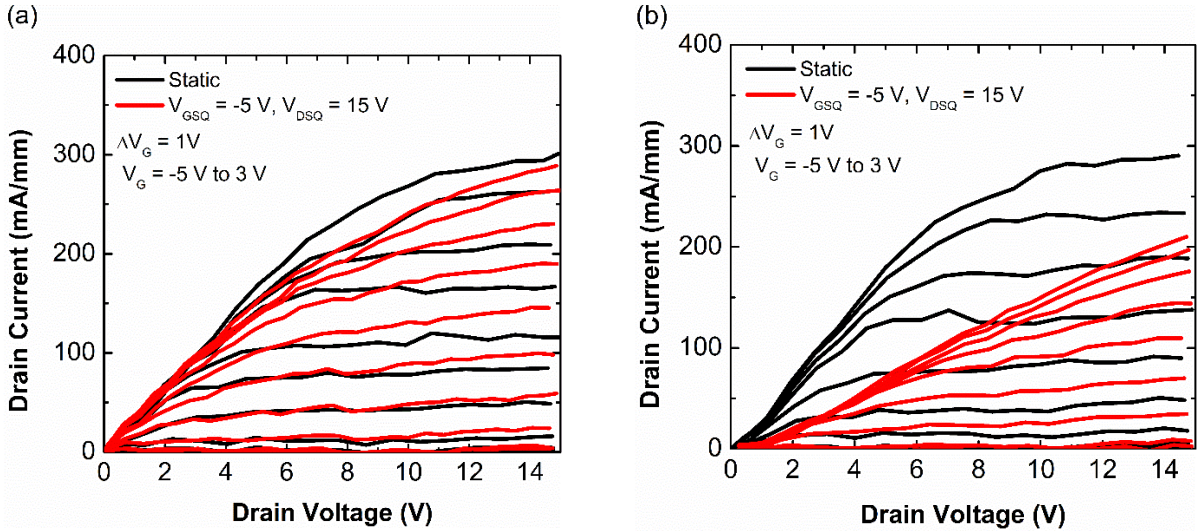

Figure 7. Pulsed I-V measurements at DC bias $V_{GS}$ = -5 ~ 3 V, $V_{DS}$ = 0 ~ 15 V (a) Sample B, (b) Sample C, each sample was measured using quiescent bias point of $V_{GSQ}$ = -5 V, $V_{DSQ}$ = 15 V with 5 µs pulse width and 5 ms period.

To assess the trap-related characteristics, pulsed I-V measurements were conducted via DIVA D265 (Figure 7). The samples showed some current collapse and knee walkout, which may be due to a either surface and buffer trap states. A higher current collapse was observed in the passivated device (Sample C), the origin of this is not clear but could be due to unintentional variation among samples during the growth.

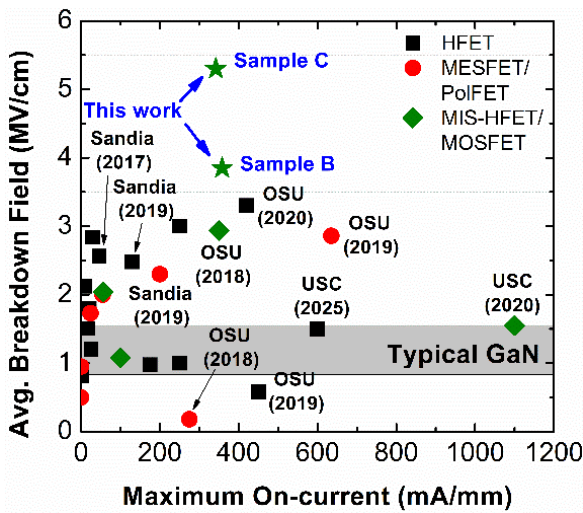

Figure 8. $F_{BR}$ vs. $I_{MAX}$ Benchmark plot of UWBG AlGaN transistors [12-32]

Finally, the fabricated devices were evaluated based on $F_{BR}$ and $I_{MAX}$, two key parameters for high-power RF applications, in comparison with previously reported UWBG AlGaN transistors. Figure 8 presents the benchmark plot of $F_{BR}$ versus $I_{MAX}$ for UWBG AlGaN transistors [12-32]. This benchmark analysis indicates that the transistors demonstrated in this study achieve state-of-the-art $F_{BR}$ with excellent $I_{MAX}$.

In conclusion, we have successfully demonstrated UWBG AlGaN metal-insulator HFETs with state-of-the-art average breakdown field and high maximum on-current. Utilizing a conventional field management approach incorporating passivation and a gate field plate, we achieved an average breakdown field of 5.3 MV/cm in $Al_{0.77}GaN/Al_{0.69}GaN$ HFET. For on-state characteristics, high maximum on-current of 358 mA/mm was realized with improved $R_{ON}$ = 3.26 Ω·mm facilitated by the implementation of a continuously grown reverse-graded ohmic contact layer. Regard RF performance, a peak $f_T$ of 9.1 GHz and $f_{MAX}$ = 15 GHz was achieved. Additionally, trap-related characteristics were assessed through pulsed I-V measurements, revealing high epitaxial quality with low gate and drain lag. These results suggest that UWBG AlGaN transistors with optimized gate dielectrics and field management techniques hold great potential for high-power mm-wave RF applications.

This work was funded by ARO DEVCOM under Grant No. W911NF2220163 (UWBG RF Center, program manager Dr. Tom Oder). This article has been authored by an employee of National Technology & Engineering Solutions of Sandia, LLC under Contract No. DE-NA0003525 with the U.S. Department of Energy (DOE). The employee owns all right, title and interest in and to the article and is solely responsible for its contents. The United States Government retains and the publisher, by accepting the article for publication, acknowledges that the United States Government retains a non-exclusive, paid-up, irrevocable, world-wide license to publish or reproduce the published form of this article or allow others to do so, for United States Government purposes. The DOE will provide public access to these results of federally sponsored research in accordance with the DOE Public Access Plan https://www.energy.gov/downloads/doe-public-access-plan

## Supplementary materials

In this material, we all included the breakdown results measured by varying the gate-to-drain spacings and the 2D TCAD simulation results to investigate the electric field distribution at breakdown condition for each sample.

## Supplementary Information

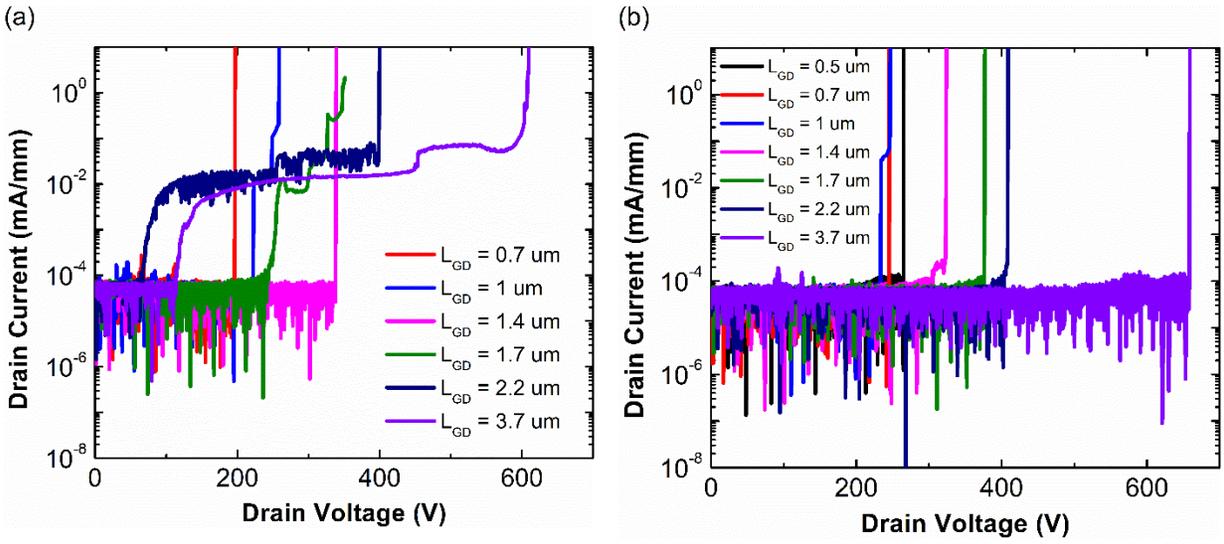

Figure S1. Three-terminal breakdown measurements of (a) Sample B varying $L_{GD}$ = 0.7 ~ 3.7 $\mu$m, (b) Sample C varying $L_{GD}$ = 0.5 ~ 3.7 $\mu$m

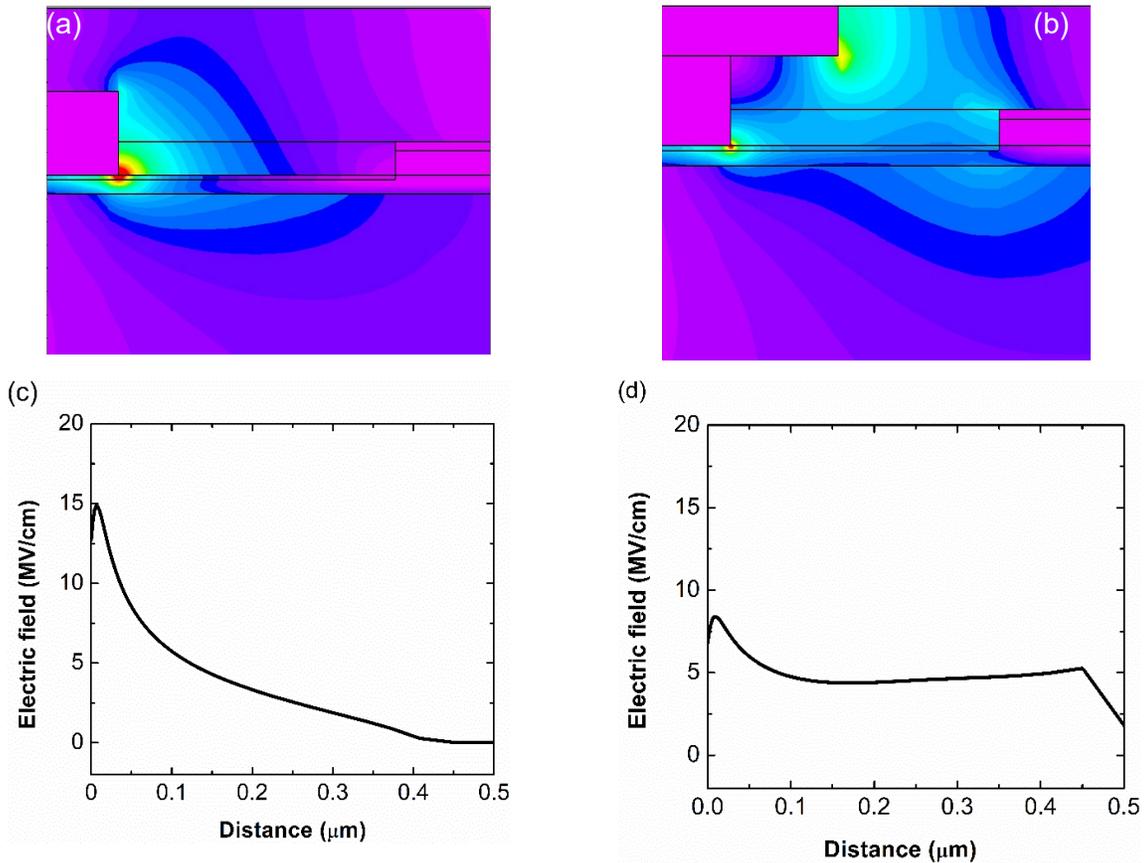

Figure S2. (a) Simulated electric field distribution of Sample B at $V_{GS}$ = -10 V and $V_{DS}$ = 193 V, (b) simulated electric field distribution of Sample C at $V_{GS}$ = -11 V and $V_{DS}$ = 260 V, (c) x-direction electric field profile of Sample B, cutline at $Al_2O_3$/AlGaN barrier interface, (d) x-direction electric field profile of Sample C, cutline at $Al_2O_3$/AlGaN barrier interface

Three-terminal breakdown measurements with varying gate-to-drain spacing are shown in Figure S1(a), (b). Some devices from Sample B show increasing leakage current with $V_{DS}$, Sample C devices (with $SiN_x$ passivation and field plates) showed almost constant leakage for all $L_{GD}$ values until the device undergoes "hard" and irreversible breakdown. Further investigation is needed to understand the exact breakdown mechanisms, but we did breakdown simulations were performed using 2D TCAD modeling [37]. From the simulation results, the x-direction peak electric field ($E_{X.PEAK}$) was evaluated as 14.8 MV/cm for Sample B under breakdown condition, while Sample C showed a significantly lower $E_{X.PEAK}$ of 8.39 MV/cm at $V_{DS}$ = 260 V. Furthermore, it was found that field-plated devices (Sample C) exhibited a more uniform, box-like electric field profile compared to non-field-plated structures (Sample B), which displayed a non-uniform field distribution. The effective peak electric field ($F_{eff}$) was calculated using $F_{OX}$ = 1.54 MV/cm, as determined in our previous dielectric study, yielding values of 14.88 MV/cm for Sample B and 8.53 MV/cm for Sample C [10]. These simulation results suggest that $SiN_x$ passivation and field plating—when applied under the same passivation thickness and field plate length conditions as the fabricated devices—can effectively reduce the peak electric field and promote a more uniform electric field distribution, thereby enabling higher voltage handling within the same device dimensions.

Differences in the leakage characteristics among devices in sample B may be due to variations in epitaxial structure, or in the process conditions/exact geometry of the gate (e.g. the real gate profile may not be a right angle as assumed in the simulation).